# The Role of Review Process Failures in Affective State Estimation: An Empirical Investigation of DEAP Dataset

Nazmun N Khan[1*], Taylor Sweet[1], Chase A Harvey[1], Calder Knapp[1], Dean J. Krusienski[2], David E Thompson[1]

[1]*Mike Wiegers Department of Electrical & Computer Engineering, Kansas State University, Manhattan, KS, USA.*
[2]*Department of Biomedical Engineering, Virginia Commonwealth University, Richmond, VA, USA.*
*1701D Platt St., Manhattan, KS 66506, USA. E-mail: nkhan1@ksu.edu

**Abstract**

The reliability of affective state estimation using EEG data is in question, given the variability in reported performance and the lack of standardized evaluation protocols. To investigate this, we reviewed 101 studies, focusing on the widely used DEAP dataset for emotion recognition. Our analysis revealed widespread methodological issues—including data leakage from improper segmentation, biased feature selection, flawed hyperparameter optimization, neglect of class imbalance, and insufficient methodological reporting. Notably, we found that nearly 87% of the reviewed papers contained one or more of these errors. Moreover, through experimental analysis, we observed that such methodological flaws can inflate the classification accuracy by up to 46%. These findings reveal fundamental gaps in standardized evaluation practices and highlight critical deficiencies in the peer review process for machine learning applications in neuroscience, emphasizing the urgent need for stricter methodological standards and evaluation protocols.

**Introduction**

Emotional human–computer interaction could be advanced by the development of affective brain-computer interfaces (aBCIs). These technologies enable the development of adaptive and automated user interfaces, as well as integration into larger systems as specialized subcomponents, enhancing user experience and interaction(*1*). To enhance the generalizability of emotion recognition algorithms across diverse subjects, researchers are increasingly focusing on developing various algorithmic approaches. These efforts include exploring different preprocessing techniques, feature extraction methods, feature selection strategies, and classifier combinations, alongside the integration of machine learning algorithms to optimize performance and accuracy. Machine learning (ML) and artificial intelligence (AI) have become integral components of modern healthcare, transitioning from experimental research to practical deployment. This shift is evident in the increasing number of AI/ML-powered medical devices receiving regulatory approval from the U.S. Food and Drug Administration (FDA)(*2*).

Although there have been significant advancements in the field of affective state estimation with the application of machine learning (*3*, *4*), we contend that the validity of these approaches is compromised by common errors and pitfalls frequently observed in aBCI experiments. The reliability of aBCI systems depends on the systematic execution of each stage within the machine learning pipeline, which includes data acquisition, noise removal, signal processing, feature extraction and selection, model selection, and performance evaluation. Common factors observed in affective state estimation using EEG that can affect performance include the mixing of training and test datasets, data leakage due to inappropriate feature selection methods, incorrect hyperparameter optimization, classifier selection, and flawed validation process. These methodological errors vary significantly in their impact, with some causing minimal degradation while others severely compromise estimation accuracy.

Segmentation, or windowing, is a widely used practice in EEG analysis, typically employed to increase the effective sample size by dividing long trials into multiple shorter segments(*5–8*). Segmenting a single trial into multiple segments is not inherently problematic. However, it is crucial to ensure that all segments from the same trial are assigned to the same data partition (training, testing, or validation). Distributing segments from a single trial across multiple partitions introduces data leakage, which can lead to significantly inflated performance estimates and compromise the generalizability of the model(*9, 10*). Another common pitfall is applying data preprocessing using global statistical properties computed from the whole dataset(*11, 12*) . For example, baseline correction, a common preprocessing step in EEG-based machine learning, can inadvertently cause data leakage if not applied correctly. Baseline correction often involves computing the mean or median EEG activity before a stimulus and subtracting it from the entire trial. If this correction is performed before splitting the data into train and test sets, it may introduce dependencies between training and test sets. Commonly utilized techniques in machine learning model development include feature/channel selection, hyperparameter optimization, and the selection of both classifiers and feature sets. However, if these processes are performed using information from the test set or the entire dataset, it can result in data leakage(*11*). Besides that, bias in emotion recognition datasets can lead to performance overestimation in machine learning models(*13*). When certain emotions are overrepresented in the data, the model may favor these emotions. Class imbalance can mislead performance metrics by favoring the majority class(*13–15*). All these issues, which are present in published studies, lead to overfitting, reduced generalizability, and misleading evaluation metrics that do not accurately reflect the real scenario of aBCI model.

Notably, the methodological challenges discussed in this context are not unique and centered only on aBCI applications(*16*). They are also commonly observed in other machine learning/ artificial intelligence applications, including healthcare, medical diagnostics, neuroimaging, data analysis, computer Security, Internet of Things (IoT) and data mining(*17–26*). Schroeder *et al.* systematically compared four cross-validation strategies to assess their impact on classifier performance estimation in passive BCI experiments (*27*). Kapoor *et al.* identified issues of data leakage and the reproducibility crisis in machine learning-based research across 17 scientific disciplines(*26*). Similarly, Balendran *et al.* (*17*) explored the impact and prevalence of various perturbations in machine learning for healthcare, including data imbalance, feature extraction and selection techniques, training strategies, and hyperparameter tuning. These factors significantly influence model robustness, generalization, and performance in medical applications. A recent review provides a comprehensive overview of how machine learning can be intentionally or unintentionally misused in brain imaging(*12*). Other previous studies have explored the impact of data bias on medical diagnostics and healthcare outcomes(*28, 16, 29, 30*). The presence of temporal bias, such as data collected at specific times or specific conditions, in current AI and robotic systems is addressed in(*31*).

Despite the promising advancements in aBCI systems and their high classification performance in controlled research environments, their transition to real-world applications remains challenging. A comprehensive analysis of aBCI research published between 2019 and 2023 revealed that only 4.58% of studies explicitly focused on online emotion classification(*13*). Another survey based on aBCI claimed that 90% of the reviewed research utilized offline classification methods, while only 8% implemented online classification, which are more applicable to real-time scenarios(*32*). This limitation is not exclusive to aBCI but reflects a broader trend in health research. The research and development community in BCI has not sufficiently prioritized the evaluation methods necessary for transitioning these systems into practical applications(*33*). Alarmingly, these pitfalls persist despite being widely recognized, raising concerns about the rigor of the peer-review process(*34–36*). The persistent approval and publication of studies that overlook these critical issues indicate a lack of diligence among reviewers.

In this study, we hypothesized that these common pitfalls occur frequently and consistently in the field of aBCI-based classification algorithms and are often inadequately addressed by reviewers during the evaluation process. Moreover, this limitation not only hinders the practical implementation of aBCI technologies but also results in inefficient use of research efforts and funding. Research inefficiency in biomedical fields represents a significant concern, with a study indicating that approximately 85% of biomedical research is largely inefficient and wasted(*37–39*). Additionally, multiple studies highlighted that these methodological flaws, lack of reproducibility, and limited applicability in real-world settings lead to the inefficient use and misallocation of healthcare research funding, ultimately restricting the contribution to scientific progress (*37, 40, 41*).

Therefore, we aimed to investigate the methodological flaws in affective state estimation as presented in the literature and to provide recommendations for avoiding these issues to enhance their reliability in real-world scenarios. To achieve this, we performed a comprehensive review of 101 emotion recognition papers using EEG, specifically focusing on the DEAP dataset(*42*), which is widely cited publicly available dataset in the field of aBCI. Subsequently, we expanded our investigation to validate the occurrence of observed pitfalls identified in the reviewed papers by conducting experimental analyses. For this purpose, we utilized both the DEAP dataset and data from EEG sensors applied to a watermelon collected in our own laboratory, inspired by the approach in (*43*). As part of this analysis, we also examined the extent of performance inflation caused by these methodological errors. Finally, we compare this performance inflation to a standardized fair analysis. This could be used as a reference for best practices to avoid these pitfalls.

**Results**

**Study selection**
Our systematic literature search, using keywords related to emotion recognition and EEG (e.g.," DEAP", "EEG emotion classification"), yielded 686 papers published between 2017 and 2023. Figure 1 illustrates a summarized overview of the selection process. After applying filters, the dataset was refined to 385 publications. At this stage, we identified 146 papers from 2023, 98 from 2022, 51 from 2021, 47 from 2020, 22 from 2019, 10 from 2018, and 11 from 2017. We implemented a two-stage selection process. In the initial stage, we selected 163 candidate papers, including the 30 papers with the highest citations from each year. For years with fewer available papers, we included all publications: 22 from 2019, 10 from 2018, and 11 from 2017. Subsequently, we prioritized full-text availability and citation metrics to select 114 papers for comprehensive review (20 papers from each year between 2023 and 2020, 18 from 2019, 8 from 2018, and 8 from 2017). During the detailed analysis phase, 13 papers were excluded due to their multimodal methodological approaches. Consequently, the final review included 101 papers for our systematic analysis and synthesis. A comprehensive list of all 101 papers is provided in Supplementary Table 1.

**Prevalence of methodological pitfalls from reviewed papers**
In our systematic review, we identified several prevalent methodological issues across the reviewed papers. Here, our analysis specifically focused on four critical methodological pitfalls that represent the most common sources of data leakage and performance inflation: segmentation strategy, feature/dimension reduction method, hyperparameter selection, and feature/model selection using grid search (Table 1). For each category, papers were categorized into one of three groups—valid, invalid, or undetermined—based on the methodological descriptions provided in the paper.

For both hyperparameter optimization and feature/dimension reduction, papers were marked as invalid when there was reason to believe—from the language in the text—that hyperparameter selection was performed by evaluating model performance on the entire dataset and dimension reduction algorithm was applied on whole dataset prior to train/test splitting. We cannot be certain whether this error actually occurred in all cases due to the lack of detailed reporting. However, the language used often led us, as secondary reviewers, to reasonably claim that the error occurred. Then, some papers mentioned performing hyperparameter search or feature reduction method, but the descriptions were so vague that it was unclear whether the process was conducted in a valid manner. These were marked as undetermined. Papers were marked as valid when there was no indication of methodological error in the text—for example, when reduction techniques were applied only to the training set, or when a dedicated validation set was used for optimization.

In the case of segmentation, papers were marked as valid when the text, figures, or reporting—such as the order of steps described in the methods section or the structure of performance matrices—indicated that segmentation was performed after train/test splitting. This also included cases where appropriate validation strategies, such as leave-one-trial-out or leave-one-subject-out, were explicitly mentioned. Papers were marked as invalid when segmentation appeared to have been conducted prior to data splitting, based on their descriptions of using k-fold or percentage-based cross-validation on already segmented data. Studies were marked as undetermined when segmentation was mentioned but the available information was insufficient to determine whether proper evaluation procedures were followed. For grid search evaluation, papers were marked as invalid if they reported a table or figure where test set performance biased the feature or model selection process.

Figure 2A presents the frequency of four common methodological pitfalls identified across 101 reviewed papers. The most frequent issue was observed in segmentation, with 58 papers marked as invalid and an additional 4 as undetermined. Hyperparameter optimization was the second prominent observed error, with 33 papers marked as invalid and 13 as undetermined—indicating that nearly half of the reviewed studies either applied questionable optimization procedures or failed to report them clearly. Feature reduction methods were also a significant concern, with 27 papers marked as invalid and 8 as undetermined. Although feature reduction methods showed 66 studies marked as valid in Figure 2A, it is important to clarify that this high number is largely due to the fact that 64 out of 101 papers did not apply any feature or dimension reduction method at all. Among the 37 papers that did apply such algorithms, only 2 papers implemented them correctly by applying the procedure solely to the training set, and were therefore marked as valid. In contrast, 27 papers applied feature/dimension reduction across the entire dataset, and 8 papers provided insufficient detail to assess their procedure. This highlights a critical issue: among studies that applied feature reduction methods, the vast majority (35 out of 37) performed so improperly or unclearly, a concern masked by the aggregated valid count. Grid search-based feature/model selection showed fewer identifiable issues, though 31 papers still exhibited methodological faults, highlighting that this area is not free from concern. To further examine the potential impact of this error, we analyzed the magnitude of inflation it could introduce. For papers that reported test set performance under multiple configurations, we collected the highest and lowest reported test accuracies—reflecting the selection of models or features based on test set results. The difference in reported test accuracy ranged from 0.24% to 49.7%. While we do not claim that all papers experienced large inflation due to these issues, the fact that accuracy could vary by up to 49% demonstrates that the impact of such methodological flaws cannot be ignored.

Figure 2B illustrates the distribution of papers based on the number of methodological faults simultaneously present, focusing on four key areas. Our analysis revealed a concerning pattern: 42 papers contained one of these methodological errors, while 33 exhibited two errors. Also, 11 studies demonstrated three concurrent methodological flaws and 2 studies presented with all four, potentially compromising their reported outcomes. Our review identified 13 studies that did not exhibit any of these four methodological errors based on their reported procedures and documentation.

A detailed evaluation of all reviewed papers across the four major methodological pitfalls is provided in Supplementary Table 2.

**Additional reporting and reproducibility concerns**
In addition to evaluating methodological procedures, we identified several additional qualities and reporting concerns. We assessed how class imbalance was addressed across the reviewed studies. Among the reviewed studies, only 38 papers reported F1 scores or complementary metrics such as precision, recall, sensitivity, and specificity, thus acknowledging potential class distribution concerns. Details are reported in Supplementary Table 2. Furthermore, we identified seven studies that reported results with confidence intervals or standard deviation bounds overlapping the theoretical chance level in at least one experimental task. Notably, these results were not acknowledged as statistically indistinguishable from chance in their respective discussions or limitations sections. This raises concerns about the rigor of peer review and highlights that studies with no statistically meaningful outcomes may still be presented—and accepted—as valid contributions.

Another concern is reporting bias, where studies selectively omit prior models with superior performance in their comparison tables. Of the 31 eligible binary classification studies (based on threshold 5 and participant-dependent analysis), 21 included comparison tables with results from previous publications (See Supplementary Table 3). Figure 2C shows the relationship between the reported performance of each paper's model and the highest accuracy listed in its own comparison table. Valence and arousal accuracies are shown separately, yielding correlations of 0.78 and 0.79 for valence and arousal tasks, respectively. Notably, only 4 out of 21 papers acknowledged literature results that exceeded their own accuracy, suggesting a tendency to underreport stronger prior models. To further quantify this bias, we identified how many higher-performing studies from our review had been published prior to each paper's submission date. Our analysis revealed that thirteen of these papers (62%) had at least three or more previously published models that achieved superior performance but were not mentioned in their comparison tables.

As part of our review, we also assessed the reproducibility of the reported studies by checking the availability of code or pretrained models. Out of the 101 papers, only eight provided accessible links to their code repositories or shared models. However, most of these links lacked sufficient content to fully replicate the reported experiments. An additional three papers stated that their code would be made available but did not include a link or any accessible resource at the time of our review.

A detailed record of all reviewed papers, including their evaluations based on each methodological criterion, is provided in the Supplementary File.

**Experimental analysis: Segmentation**

In this experiment, we performed binary classification on the valence axis of the DEAP dataset, after segmenting each trial into multiple segments. Additionally, binary classification was conducted on the watermelon dataset with randomly generated binary labels. Figure 3 shows how different validation strategies impact classification performance when segmentation or windowing is applied, using the DEAP dataset and watermelon dataset. The graph highlights a notable difference between two validation methods: k-fold cross-validation applied to all segmented trials, and leave-one-trial-out, which is applied to each original trial with all segmented trials in one-fold, as the number of segments per trial increases. In the Figure 3A, the k-fold cross-validation method shows a significant improvement in performance, increasing from about 53% accuracy without segmentation to over 90% with just 6 segments per trial, and nearly 100% with 60 segments. On the other hand, the leave-one-trial-out validation method remains relatively consistent at around 52% accuracy, regardless of the segmentation level. A comparable phenomenon was also observed in the binary classification experiments using the watermelon dataset, where similar patterns of performance inflation occurred despite the absence of meaningful neurophysiological signals (Fig. 3B). These results conclusively demonstrate that conventional k-fold cross-validation after segmentation can cause inflated performance estimates.

**Experimental analysis: Feature/dimension reduction method**

To assess the performance inflation caused by improperly implemented feature reduction procedures, we conducted an experimental comparison of valid and invalid reduction strategies using the DEAP and Watermelon datasets (Table 2). To perform binary classification, we tested two validation approaches. The Global Selection approach—where feature selection is applied on the whole dataset—consistently yielded inflated accuracy compared to the Local Selection approach which exclusively used training data for feature selection. Two performance metrics are reported: regular accuracy, which does not account for class imbalance, and balanced accuracy, which accounts for class bias. For the DEAP dataset in the valence axis, Global Selection inflated conventional accuracy by 15.2 percentage points (77.4% versus 62.2%) and balanced accuracy by 15.9 percentage points (75.3% versus 59.4%). Similar patterns were observed for arousal (inflation of 14.2 and 16.2 percentage points) and dominance (inflation of 15.9 and 18.1 percentage points) dimensions. Notably, the watermelon dataset—containing no neurophysiological signals—exhibited a 17-percentage point inflation in regular accuracy when using the Global Selection methodology. Since this dataset contains balanced class labels, regular accuracy and balanced accuracy are almost equivalent by definition; thus, only regular accuracy is reported. These findings quantitatively demonstrate that improper implementation of feature reduction algorithms-particularly the use of test-set data during feature ranking-can substantially overestimate classification performance, with inflation magnitudes ranging from approximately 14-18 percentage points across all conditions.

**Experimental analysis: Hyperparameter optimization**

In our hyperparameter optimization experiments, we observed consistent performance disparities between compromised and methodologically correct approaches (Table 3). Two experimental approaches were employed: (i) hyperparameter optimization applied to both the training and test sets, and (ii) optimization restricted to the training set only. In the valence axis, the regular accuracy increased by 21.9 percentage points when hyperparameter tuning was improperly performed using test data (84.4% vs. 62.5%). Similar inflations were observed for arousal (21.6%) and dominance (21.4%). The balanced accuracy metric, which compensates for class imbalance, also showed substantial inflation: 24.2, 25.2, and 24.9 percentage points for valence, arousal, and dominance, respectively. Most strikingly, our watermelon dataset showed an increase in accuracy from 50.5% to 77.6%—a 27.1 percentage point inflation—when hyperparameters were improperly optimized using test data. These results highlight that improper inclusion of the test set in hyperparameter optimization leads to an overestimation of performance, with inflation magnitudes ranging from approximately 21 to 27 percentage points across all evaluated conditions.

**Discussion**

In this study, we investigated the prevalent methodological errors commonly encountered in aBCI experiments by analyzing the most highly cited publications utilizing the DEAP dataset from 2017 to 2023. Our findings indicate that while many studies report high-performance results in EEG-based emotion estimation, a significant portion of these results may be attributed to the methodological pitfalls identified in DEAP-related research. The high prevalence of data leakage during segmentation (58 publications) highlights a fundamental misunderstanding of proper validation protocols. Similarly, the misapplication of feature reduction algorithm (27 publications) and hyperparameter optimization (33 publications) across entire datasets rather than restricting them to the training phase indicates a critical methodological gap. Notably, almost 87% of the reviewed papers exhibited at least one or more of these four pitfalls. In addition to these explicit errors, we identified a substantial number of papers where the methodological description was insufficient to assess validity —specifically, 4 in segmentation, 8 in feature reduction, and 13 in hyperparameter optimization. Many of these studies failed to document critical methodological details, such as thresholding criteria, data partitioning strategies, or whether analyses were subject-dependent. Although these cases were not labeled as definitively invalid, these undetermined categories also reflect a broader concern with inadequate transparency. Vague or incomplete methodological descriptions obscure potential errors that should be identified during the peer review process.

This concern is further underscored by additional reporting issues we observed, including limited reporting of class-imbalance metrics, failure to acknowledge results statistically indistinguishable from chance, and a lack of acknowledgment of prior models that achieved higher accuracy. We found that even within the limited scope of this review, authors seem to be selectively choosing papers with lower performance than their own. Given that the primary contribution of these papers is often the model performance itself, failing to cite models with higher performance is a major issue. Together, these findings highlight not only pervasive implementation errors, but also a lack of reporting rigor and review scrutiny.

Our experimental analysis demonstrated that high classification accuracy on DEAP dataset can be achieved when single-trial segmentation, especially when train-test data leakage occurs due to improper validation strategies. The gap between the two validation methods—about 45 percentage points at the highest segmentation—shows how improper validation can result in misleading high-performance metrics in affective BCI research. Moreover, hyperparameter optimization or feature reduction errors in individual can inflate performance estimates by approximately 14-25%. More concerning is that multiple methodological errors can occur simultaneously, which could significantly inflate the classification accuracy. The inflated results from the watermelon dataset confirm that improper application of segmentation, dimensionality reduction, or hyperparameter optimization can substantially inflate reported model performance—even in the absence of meaningful neural signals. In this experimental study, we conducted an in-depth analysis of the segmentation, feature/dimension reduction method, and hyperparameter search; however, the effects of bias and data imbalance were not extensively examined. For in-depth observation, we refer to our previous analysis which demonstrated that neglecting class imbalance can lead to misleading results(*15*).

Our findings also highlight that a lack of expertise in machine learning can lead to overestimation, raising concerns about the reliability of current aBCI research. Notably, our review was conducted solely based on the documented methodologies and reported procedures in the publications. While only 8 out of 101 papers provided accessible code repositories, we examined available code when provided to verify methodological descriptions. However, the majority of our assessment relied on the information reported in the publications themselves, without access to training models or implementation details. The identification of these pitfalls through careful examination of published materials indicates that these issues should be detectable during peer review. Our ability to identify these methodological flaws using only the information available in publications suggests the potential gaps in the rigor of the review process, where critical methodological details may not receive sufficient scrutiny. This finding underscores the need for more thorough methodological assessment during peer review to ensure research validity and reproducibility.

Therefore, we propose some critical recommendations to mitigate the pitfalls observed in our review. A primary concern is preventing data leakage between training and testing datasets, as such contamination invalidates the reported performance. To ensure a clear separation, all methodological procedures must remain independent of test data(*26*). All methodological decisions—including preprocessing techniques, feature reduction approaches, and hyperparameter optimization—should be executed exclusively on training data, without incorporating information from the test set(*11*, *14*). Decisions regarding classifier selection and feature type, and frequency bands, should derive solely from training data performance rather than prioritizing combinations that maximize test set outcomes. Choosing a suitable validation strategy is essential, especially for windowing/segmentation. Leave-one-trial-out or leave-one-subject-out are more appropriate. Our experimental analysis also demonstrates the correct application of these approaches, serving as a guideline to avoid common pitfalls. The code is available upon request. Additionally, appropriate performance metric selection is essential for addressing class imbalance issues; conventional metrics such as F1 score, precision, and sensitivity may prove inadequate when calculated predominantly on majority class samples. Balanced accuracy represents a better alternative metric due to its inherent insensitivity to class imbalance(*15*). To ensure the model robustness, it should be evaluated on multiple datasets and through online testing. Researchers must ensure that the test set is unbiased and representative of the target distribution. Furthermore, insufficient methodological details hinder model reproducibility. All model parameters, random seeds, and relevant information should be clearly reported. Additionally, all code, training models, and datasets should be accessible to all. Finally, ensuring that reviewers have comprehensive knowledge across relevant domains is crucial. Insufficient knowledge in any single aspect of BCIs or machine learning can lead to deficiencies in experimental design, statistical analyses, or result interpretation. Adding reviewers from multidisciplinary field might be helpful to identify methodological shortcomings in both fields. Additionally, reviewers should critically evaluate model reproducibility and availability during the review process.

Our review methodology has certain limitations. We utilized only the Scopus database for our literature search, potentially limiting the scope of included studies from multi-database approaches. The keyword search specifically targeted studies on emotion recognition using the DEAP dataset, selected due to its high citation frequency in the field. These limitations may have led to the exclusion of broader developments in the field. However, we believe that using a single dataset allows for a more consistent and comparable evaluation of methodological errors. Future reviews would benefit from incorporating multiple databases, expanding the study period, and including studies utilizing diverse emotional datasets to provide more comprehensive insights. Determining whether a method within a paper committed a methodological error based solely on the text is a difficult task, and inherently asks the reviewer to make many judgement calls.

To reduce bias, multiple reviewers assessed each paper and resolved discrepancies through discussion. While individual assessments may not be flawless, we believe this approach provides a valid evaluation of the literature and supports the reliability of our aggregate findings.

In conclusion, our comprehensive analysis of the affective BCI literature reveals widespread methodological shortcomings that significantly impact the field's scientific rigor. By identifying and replicating these issues, we emphasize the necessity for stricter validation strategies, transparent reporting, and unbiased evaluation metrics. Our findings are not merely a critique but a call to the scientific community to uphold higher standards in experimental design, peer review, and result interpretation. Failure to do so may compromise the integrity of the findings and continue to perpetuate biased and misleading studies. Addressing these concerns will enhance the reliability of aBCI research, fostering reproducible and impactful advancements in the field.

## Materials and Methods

### Review paper Strategy

*Search strategy and selection criteria*: In this study, we used the Scopus database to identify relevant studies for our review. Our search criteria were designed to be selective, focusing on core research papers central to our topic as of February 29, 2024. We specifically targeted affective state estimation studies that used the DEAP dataset, as DEAP is one of the most widely cited databases in emotion recognition(*44*). Therefore, we focused on "DEAP" as a keyword in our search. Using keyword combinations such as "EEG," "emotion," "recognition," "using," and "DEAP," we conducted an initial search (Fig. 1). We limited the publication years to 2017 through 2023. Afterward, we applied filters to restrict the language to "English," document type to "article," and source type to "journal". These filtering criteria were implemented to focus solely on journal publications, which typically provide more comprehensive methodological details and novel applications compared to other publication formats. The collected publications were then organized for each year, by ranking them from highest to lowest citations. To examine the methodological developments over time, we initially aimed to select the 30 most highly cited papers from each year between 2023 and 2020, while including all papers from 2019, 2018, and 2017 due to their limited numbers. Then, we attempted to select an equal number of 20 papers from each year by considering the full article availability and citation metrics. Finally, we manually excluded the studies that applied multimodal emotion recognition to restrict the focus to EEG-based affective state estimation.

*Review process and data extraction*: The paper review process was conducted in four stages to ensure rigor and transparency. In the first stage, the initial paper search and selection were performed by one author based on the criteria mentioned above. In the second stage, each selected paper was independently assessed by two authors to evaluate its methodological quality and pitfalls. In the third stage, any discrepancies or conflicts in the assessments were resolved through group discussions involving all authors. Finally, in the fourth stage, the reported outcomes were systematically re-evaluated by all authors to ensure clarity and consistency.

*Methodological quality assessment criteria:* Based on the extracted information from the revision, the papers were then categorized based on four common methodological pitfalls: data leakage due to segmentation, data leakage in hyperparameter optimization, data leakage in feature/dimension reduction methods, and feature/model selection using grid search. The defined concepts for each of these issues are detailed in Table 1. For each identified methodological pitfall, papers were systematically classified into three categories based on the clarity and appropriateness of their reported procedures:

- Valid: Papers that clearly described proper methodological procedures, with explicit indications of appropriate data partitioning and no indication of data leakage or improper practices.
- Invalid: Papers where the described methodology indicated violations of best practices, data leakage or procedures that would compromise model generalizability.
- Undetermined: Papers that mentioned relevant procedures but provided insufficient detail to determine whether the approach followed a valid evaluation process, often due to vague or incomplete reporting.

*Additional reporting quality indicators:* We evaluated additional quality indicators including: (1) class imbalance acknowledgment, (2) statistical significance relative to chance performance, and (3) code availability for reproducibility.

For class imbalance assessment, papers were classified as "reported" if they included any of the following metrics that account for class distribution: F1 scores, precision, recall, sensitivity, or specificity. Papers reporting only accuracy or similar aggregate metrics were marked as "not reported". For statistical significance evaluation, we identified studies reporting confidence intervals or standard deviation bounds and determined whether these overlapped with theoretical chance levels. Theoretical chance levels were calculated based on the number of classes in each classification task (50% for binary classification, 25% for four-class classification). Studies were flagged when the chance classification level was within two of the reported standard deviations, indicating results potentially indistinguishable from random performance. For code availability assessment, papers were marked as "available" only if they provided accessible links to code repositories or pretrained models. Studies that mentioned code availability but lacked any accessible links or provided only pseudocode or implementation summaries within the paper were not counted.

*Literature comparison bias analysis:* To assess selective reporting bias, we investigated the extent to which published studies accurately report prior models with superior performance. To ensure fair comparison, we restricted our analysis to papers that performed participant-dependent binary classification using a threshold of 5 on the valence or arousal dimension—yielding a subset of 31 papers from our original collection. Of these, 21 papers included explicit comparison tables listing prior models. For each of these 21 papers, we extracted: (1) the reported accuracy of each paper's proposed model, and (2) the highest accuracy reported in their comparison table from prior studies. Pearson correlation coefficients were computed to assess the relationship between proposed and literature-reported accuracies for valence and arousal tasks separately. To further quantify underreporting, we cross-referenced our review database to identify how many high-performing models had been published prior to each paper's submission date.

**Emotion dataset DEAP**

The DEAP dataset(*42*) is a widely utilized benchmark for emotion recognition using EEG signals. It consists of EEG recordings from 32 healthy participants (50% female, aged 19–37; mean age: 26.9 ± 4.45) while they watched 40 music videos designed to elicit emotional responses. The database provides two datasets: raw data and preprocessed data. The raw data includes 32 BioSemi .bdf files, each containing 48 recorded channels: 32 EEG channels, 12 peripheral channels (including electrooculography (EOG), electromyography (EMG), galvanic skin response (GSR), and respiration), three unused channels, and one status channel. The EEG signals were originally recorded at a sampling rate of 512 Hz. To facilitate analysis, a preprocessed version of the dataset was provided, incorporating several modifications. The data was downsampled to 128 Hz, EOG artifacts were removed, and a 4–45 Hz band-pass filter was applied to retain relevant EEG frequency components. Additionally, the signals were re-referenced using a common average reference, and the data was segmented into 60-second trials, with a 3-second pre-trial baseline.

Each participant has 40 trials. Following each video stimulus, participants provided self-assessment ratings on a nine-point scale for four emotional dimensions: valence, arousal, dominance, and liking. Emotional states in the DEAP are often interpreted using Russell's circumplex model, where valence and arousal define a two-dimensional emotional space.

**Watermelon dataset**

The EEG collection method using watermelon was previously referred to as phantom EEG in earlier studies(*43*, *45*, *46*). The watermelon EEG data focuses solely on temporal correlation features without capturing any stimulus-driven neural responses(*43*). Here, watermelon EEG data acquisition was performed using the Cognionics Mobile-72 EEG system, a high-density mobile EEG device with 64 channels (Fig. 4A). We collected EEG data from the watermelon for more than one hour to ensure sufficient data for the classification task. The recordings were conducted at a sampling frequency of 500 Hz. The EEG cap, equipped with standard active Ag/AgCl electrodes, was positioned according to the international 10–20 system, a widely recognized standard in EEG studies. Reference and ground electrodes were placed on either side of the watermelon to mimic standard mastoid placements and ensure signal quality. The watermelon EEG data was then reorganized to follow the same structure as the DEAP dataset. From the first one-hour recording, 40 trials were extracted, each lasting 60 seconds. A 30-second interval was maintained between consecutive trials to avoid unwanted influences between trials. These 40 trials were then randomly assigned binary labels, either 1 or 0, for the final classification task.

**Procedure of experimental analysis**

In a typical EEG-based emotion recognition classification framework, the process generally follows a standardized pipeline that includes data preprocessing, feature extraction, feature selection, train-test partitioning, classifier selection, and performance evaluation using metrics such as accuracy and F1-score. In this study, we implemented all these steps; however, to systematically assess the significance and impact of specific methodological pitfalls, we prioritized one pitfall per experiment. Accordingly, we conducted separate experiments focusing on segmentation, feature selection, and hyperparameter optimization. In all cases, the same experimental protocol was applied consistently to both the DEAP and watermelon datasets.

**Segmentation analysis**

To investigate the improper use of EEG trial segmentation, we performed subject-dependent binary classification using the raw DEAP dataset. Each trial in the DEAP dataset originally lasted 60 seconds (Fig. 4B). To increase the sample size, we segmented each trial into t-second intervals, resulting in a total of $N=60/t$ segments per trial. For instance, segmenting into 1-second intervals resulted in a total of 2400 trials (40 trials × 60 seconds). By varying the segment length ($t$), we assessed the effect of segmentation on classification performance. A threshold of 5 was applied, where ratings less than or equal to 5 (≤5) were considered low states, and ratings greater than 5 (>5) were considered high states. For each emotional axis, there are two classes: high valence vs. low valence, high arousal vs. low arousal, and high dominance vs. low dominance. In emotion recognition studies, EEG data are often analyzed across five frequency bands—delta (1-4 Hz), theta (4-8 Hz), alpha (8-14 Hz), beta (14-30 Hz), and gamma (30-42 Hz). The power spectral density (PSD), one of the most used features in aBCI classification, was computed using MATLAB built-in 'pwelch' function for those frequency bands as the feature for classification.

Various classification algorithms, including decision trees, logistic regression, Naive Bayes, k-nearest neighbors (kNN), support vector machine (SVM), CNN, and neural networks (NN) have been employed for binary classification in prior studies. In our study, we selected kNN as the primary classifier for categorizing binary emotional states across all three dimensions. The kNN classifier was implemented using MATLAB's built-in 'fitcknn' function with default parameters, with the number of nearest neighbors (k) specifically set to 5. To rigorously evaluate the classifier's performance, we performed two distinct validation methodologies: k-fold cross-validation and leave-one-trial-out validation. For the former approach, 5-fold cross-validation was applied to all 2400 samples from each individual subject. In the latter approach, we implemented a leave-one-trial-out validation method across the original 40 trials, ensuring that all segmented samples derived from a single trial remained within the same fold. This methodological choice was crucial to prevent potential data leakage between training and testing sets that could occur due to segmentation. Specifically, the training data comprised segmented samples from 39 trials, while the testing data was exclusively derived from the remaining trial. This process was iteratively repeated until each trial had served as a test set. Here, a well-known performance metric, balanced accuracy, was also applied to evaluate the performance of both validation methods. Balanced accuracy, which accounts for the potential class imbalance in the emotional state classification task, is calculated by averaging the individual class accuracy for all classes(*15*). The same preprocessing protocols, feature extraction techniques, and classification methodology were subsequently applied to the watermelon dataset for comparative analysis.

**Feature selection analysis**
Here, we employed the DEAP preprocessed dataset which is also widely used for affective state estimation research. We analyzed all trials of 60-second experimental signals after removing the 3-second baseline, or and applied the same thresholding with a value of 5 to divide the 40 trials into binary classes. As the DEAP preprocessed data was already filtered through a 4-45 Hz band-pass filter, we calculated the PSD across four frequency bands: theta (4-8 Hz), alpha (8-14 Hz), beta (14-30 Hz), and gamma (30-42 Hz). For feature selection, we implemented a statistical t-test using MATLAB's built-in "rankfeatures" function to identify the most relevant features while reducing redundancy. Classification was performed using the kNN classifier with default parameters. However, we conducted subject-dependent experiments using 70% of the trials as the train set and 30% of the trials as the test set. We systematically performed two feature selection approaches. Two feature selection strategies were compared: (1) Global Selection, where t-test feature selection was applied to the entire dataset before splitting into train and test sets—a common but flawed practice that can introduce data leakage by allowing test data to influence feature selection, potentially inflating performance metrics; and (2) Local Selection, where the training set underwent 5-fold cross-validation. In each cross-validation fold, t-test feature selection was applied using only the training data (i.e., 4 out of 5 folds), ensuring that the validation fold remained completely unseen during feature selection. The selected top-ranked features were then used to train a kNN classifier on the training data, and the model was evaluated on the remaining 1-fold validation set. This was repeated for all folds and across a range of feature counts. The average classification accuracy across the five folds was computed for each feature count, and the number of features yielding the best performance was selected as optimal. After determining the optimal count, the final feature set was re-ranked using the full training set and applied to the independent test set for final evaluation. The latter approach maintains strict separation between training and test data, preventing information leakage and yielding more realistic performance estimates despite potentially introducing variability in selected features across folds. The effectiveness of both approaches was assessed using both accuracy and balanced accuracy as mentioned above.

**Hyperparameter optimization**
In this analysis, we also employed the preprocessed DEAP dataset, maintained identical pre-processing and feature extraction methodologies as outlined in the preceding section. We also applied the same validation approach, a subject-dependent experiments using a 70/30 train/test split. To investigate potential data leakage during hyperparameter optimization, we implemented two distinct approaches with a kNN classifier. In the first approach, we repeatedly adjusted the classifier's hyperparameters (number of neighbors, standardization setting, and distance metric), selecting the combination that yielded the highest classification accuracy on the test data. This method, while yielding optimal performance metrics, risks overfitting to the test set. In contrast, the second strategy followed proper machine learning practice by isolating the test set entirely during the tuning process. We employed a 5-fold cross-validation procedure within the training set to optimize the combination of hyperparameters, including the number of neighbors, standardization (enabled or disabled), and type of distance metric. For each hyperparameter setting, the training data was partitioned into five folds. In each fold, the model was trained on 4 folds and validated on the remaining one. The balanced accuracy was computed on the validation fold and averaged across all five folds. The combination of hyperparameters that achieved the highest average balanced accuracy was selected as optimal. The final model was then retrained on the full training set using these optimal settings and subsequently evaluated on the untouched test set. This approach ensured that no test data was used during model tuning, thus avoiding data leakage and providing a more reliable estimate of generalization performance. Both accuracy and balanced accuracy were chosen as the performance metrics to compare both approaches, providing insights into how optimization methodology impacts classification results.

**Figures and Tables**

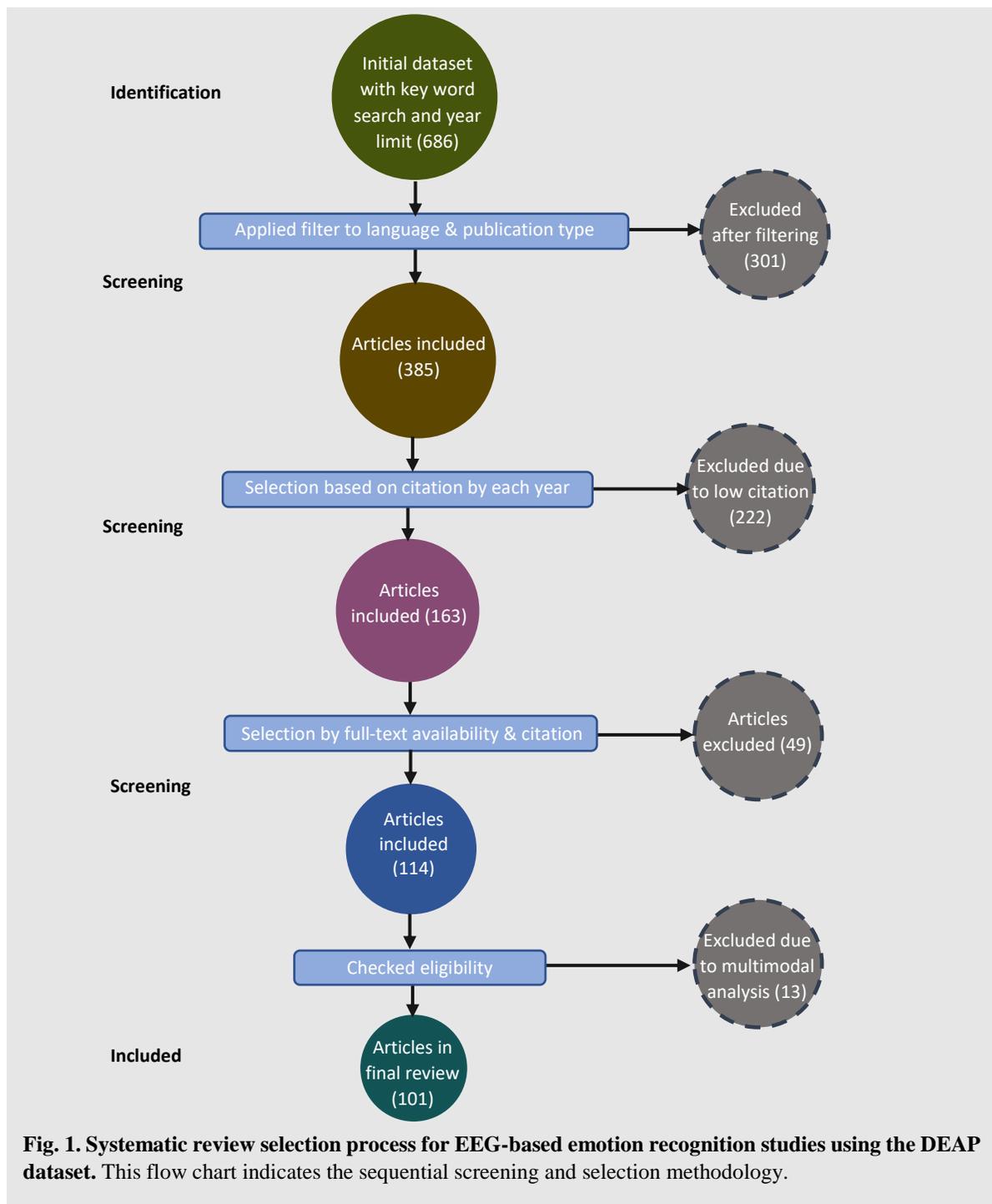

**Fig. 1. Systematic review selection process for EEG-based emotion recognition studies using the DEAP dataset.** This flow chart indicates the sequential screening and selection methodology.

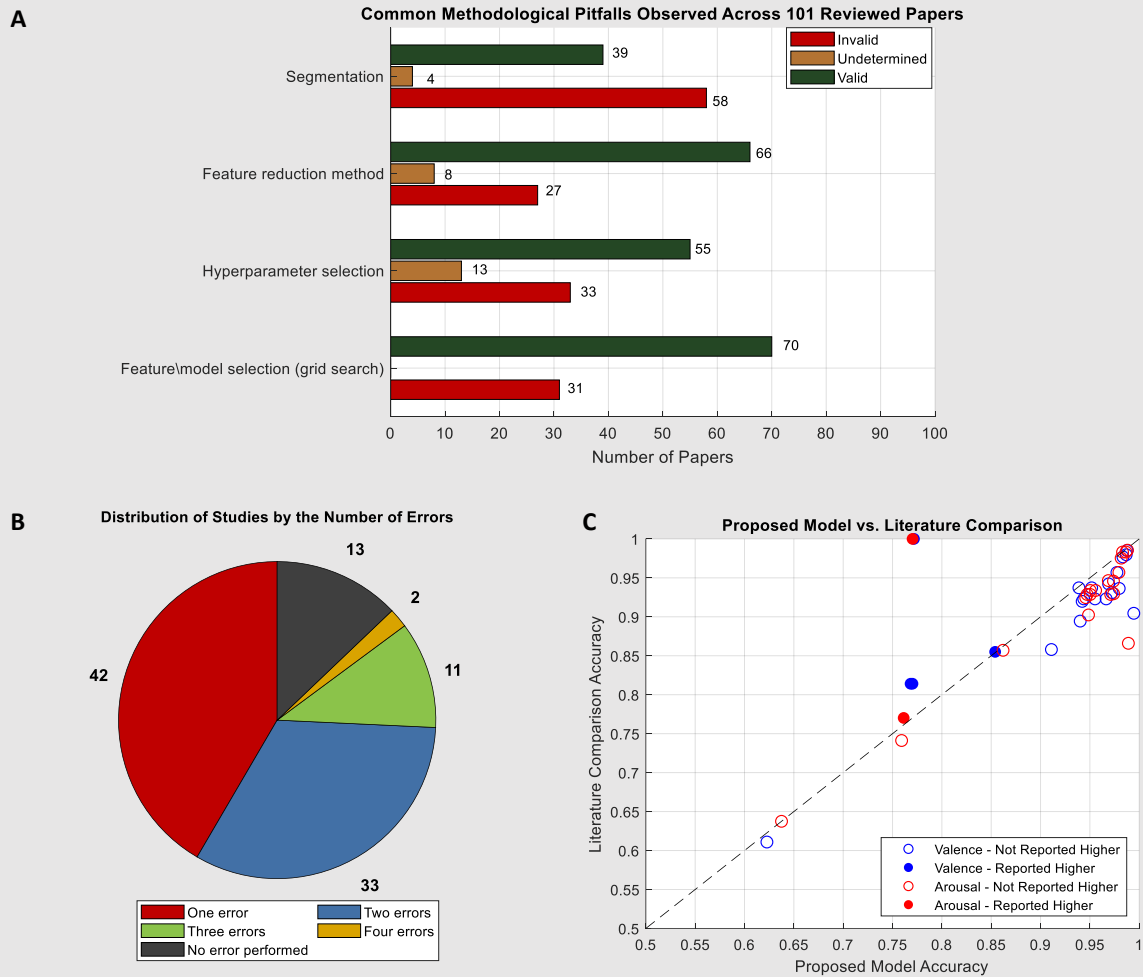

**Fig. 2. Prevalence and distribution of methodological errors in EEG-based emotion recognition studies.** (**A**) Distribution of common methodological issues across 101 reviewed studies, focusing on segmentation, feature reduction method, hyperparameter tuning, and feature/model selection using grid search. Studies were categorized as valid (green), undetermined (yellow), or invalid (red). (**B**) Distribution of studies based on the number of methodological errors identified, showing the proportion of papers containing one error (n=42), two errors (n=33), three errors (n=11), four errors (n=2) and no errors (n=13). (**C**) Comparison of proposed model accuracy versus the highest prior accuracy cited in 21 studies. Blue and red circles represent valence and arousal tasks, respectively; filled markers indicate papers that acknowledged higher-performing prior models. Correlations: r = 0.78 (valence) and r = 0.79 (arousal).

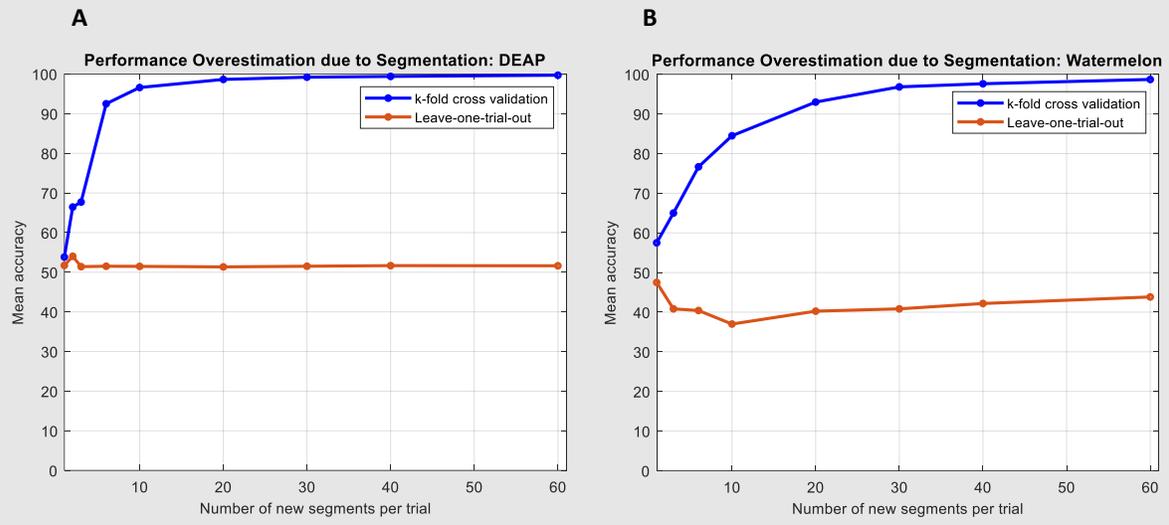

**Fig.3. Performance overestimation due to windowing/segmenting trial in EEG-based binary classification.** (**A**) Comparison of mean accuracy between k-fold cross validation and leave-one-trial-out validation approaches as segments per trial increase using DEAP dataset in the valence axis. (**B**) Binary classification accuracy trends for watermelon dataset under the same segmentation conditions. In both cases, k-fold cross validation (blue) results in inflated accuracy estimated compared to leave-one-trial-out validation (orange).

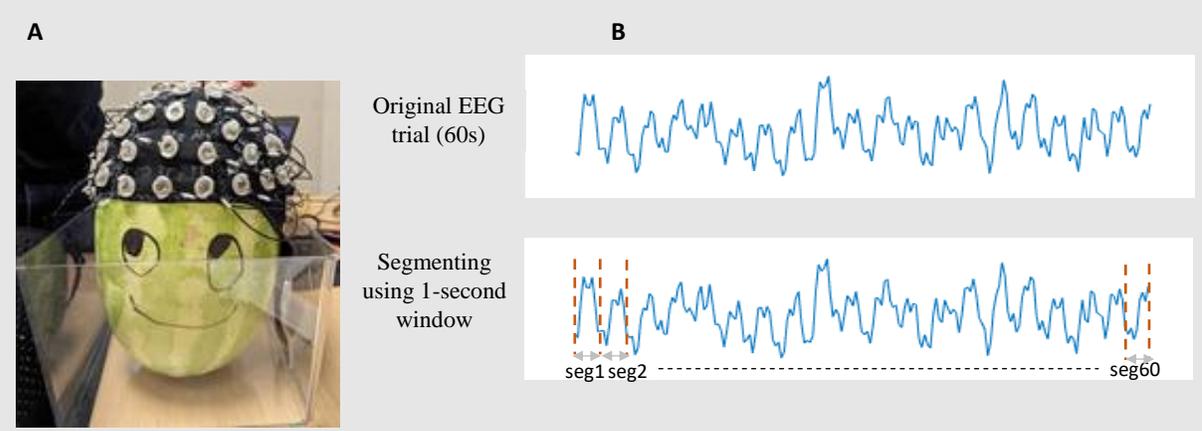

**Fig.4. EEG data collection, and segmentation.** (**A**) Demonstration of EEG data collection using a watermelon phantom. (**B**) EEG trial segmentation process, illustrating a 60-second EEG trial divided into 1-second non-overlapping windows, resulting in 60 segments per trial. This approach is commonly used to augment sample size but may introduce data leakage if segments from the same trial are distributed across training and test sets. Segments are illustrative and not shown to scale.

**Table 1. Definitions and examples of common methodological pitfalls identified in EEG-based emotion recognition studies.** Four major categories of methodological issues were used to assess the reviewed literature: data leakage due to segmentation, hyperparameter tuning bias, improper application of feature or dimension reduction, and feature/model selection using grid search. The table provides a brief definition and representative example for each pitfall category used in the review.

| Pitfalls | Definitions | Examples |
|---|---|---|
| Data leakage due to segmentation | Presence of samples which are derived from the same trial after segmentation or windowing, in both the training and testing sets. | A 1-second segmentation was applied to each 60-second trial, resulting in 2400 segments from 40 trials. A k-fold cross-validation was then performed on these 2400 segments, revealing the presence of samples from the same trial in both the training and testing sets(*47*). |
| Bias in hyperparameter tuning | Hyperparameter optimization based on accuracy of both train and test sets. | Selected configuration and settings of the classifier based on highest classifier performance(*48*). |
| Data leakage due to feature/dimension reduction method | Dimensionality or feature reduction algorithm was performed using the entire dataset before splitting into training and test sets, which can lead to data leakage and inflated performance estimates. | Dimension reduction applied on whole dataset before train/test set partition(*49*). |
| Feature/model selection using grid search | Feature or model selection was conducted using grid search procedures where the selection criteria were based on performance metrics evaluated on the test set. | 1. Selected the best electrodes based on highest performance accuracy(*48*). 2. Selected classifier based on highest test set accuracy(*49*). |

**Table 2. Comparison of Feature selection strategies in binary classification across DEAP and watermelon datasets.** Binary classification accuracy is reported for DEAP and Watermelon datasets using Global and Local feature selection approaches. For the Watermelon dataset, which uses randomly generated balanced labels, only regular accuracy is reported, as it is equivalent to balanced accuracy.

|  |  | **Regular accuracy** | | **Balanced accuracy** | |
|---|---|---|---|---|---|
|  |  | Global | Local | Global | Local |
| **DEAP** | Valence | 77.41 | 62.20 | 75.31 | 59.39 |
|  | Arousal | 76.77 | 62.53 | 70.39 | 54.17 |
|  | Dominance | 77.17 | 61.24 | 71.63 | 53.54 |
| **Watermelon** |  | 64.00 | 47.00 | -- | -- |

**Table 3. Comparison of Hyperparameter Optimization Strategies in Binary Classification.**
Binary classification performance is reported for DEAP and Watermelon datasets using two hyperparameter tuning strategies: incorrect (based on test set performance) and correct (restricted to training data only). Across all conditions, using improper optimization led to substantially inflated accuracy—by over 20 percentage points in some cases.

|  |  | Regular accuracy | | Balanced accuracy | |
|---|---|---|---|---|---|
|  |  | Wrong | Correct | Wrong | Correct |
| **DEAP** | Valence | 84.42 | 62.52 | 83.16 | 58.91 |
|  | Arousal | 82.26 | 60.69 | 78.95 | 53.79 |
|  | Dominance | 82.96 | 61.56 | 78.01 | 53.04 |
| **Watermelon** |  | 77.66 | 50.50 | -- | -- |